\documentclass[aps,showpacs,amssymb]{revtex4}
\usepackage{graphicx}
\usepackage{amsfonts}

\newcommand{\beq}{\begin{equation}}
\newcommand{\eeq}[1]{\label{#1} \end{equation}}

\newcommand{\bed}{\begin{displaymath}}
\newcommand{\eed}{\end{displaymath}}
\def\bea{\begin{eqnarray}}
\def\eea{\end{eqnarray}}

\begin{document}
\date{\today}

\title{A Kinematically Complete Analysis of the CLAS Data on the
Proton Structure Function $F_2$ in a Regge-Dual Model}

\author{R.~Fiore$^{a}$,
A.~Flachi$^{b}$,
L.L.~Jenkovszky$^{c}$,
A.I.~Lengyel$^{d}$,
and V.K.~Magas$^{c,e}$
}

\address{$^{a}$ Dipartimento di Fisica, Universit\`a della
Calabria $\&$ Instituto Nazionale di Fisica Nucleare, Gruppo collegato
di Cosenza, I--87036 Arcavacata di Rende, Cosenza,
Italy} \vskip 0.2cm
\address{$^b$ IFAE, Campus UAB, 08193 Bellaterra (Barcelona), Spain}
\vskip 0.2cm
\address{$^{c}$ Bogolyubov Institute for Theoretical Physics, Academy of Sciences
of Ukraine, UA-03143 Kiev, Ukraine} \vskip 0.2cm
\address{$^{d}$ Institute of Electron Physics, Universitetska 21, UA-88000
Uzhgorod, Ukraine} \vskip 0.2cm
\address{$^e$ Center for Physics of Fundamental Interactions (CFIF),}
\address{Physics Department, Instituto Superior Tecnico, Av. Rovisco
Pais, 1049-001 Lisbon, Portugal}

\begin{abstract}
The recently measured inclusive electron-proton cross section in
the nucleon resonance region, performed with the CLAS detector at
the Thomas Jefferson Laboratory, has provided new data for the
nucleon structure function $F_2$ with previously unavailable
precision. In this paper we propose a description of these
experimental data based on a Regge-dual model for $F_2$. The basic
inputs in the model are nonlinear complex Regge trajectories
producing both isobar resonances and a smooth background. The
model is tested against the experimental data, and the $Q^2-$
dependence of the moments is calculated. The fitted model for the
structure function (inclusive cross section) is a limiting case of
the more general scattering amplitude equally applicable to deeply
virtual Compton scattering (DVCS). The connection between the two
is discussed.\\
\end{abstract}



\maketitle

\section {Introduction} \label{s1}

It has been recently realized \cite{DVCS,R1,R2,R5} that a
straightforward generalization of the ordinary parton densities
arises in exclusive two-photon processes in the so-called
generalized Bjorken region, e.g. in Compton scattering with a
highly virtual incoming photon, and in the hard photoproduction of
mesons. Here one finds off-forward matrix elements, as
distinguished from the forward ones in inclusive reactions.

\begin{figure}[htb]
\begin{center}
\includegraphics[height=6cm,angle=0]{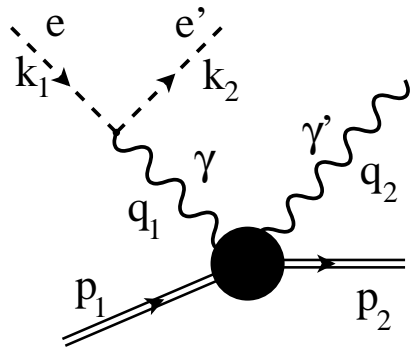}
\end{center}
\caption{Kinematic of deeply virtual Compton scattering.}
\label{dvcs}
\end{figure}

Deeply virtual Compton scattering (DVCS) combines the features of
the inelastic processes with those of an elastic process. The
diagram of such a process, $e(k_1)+p_1 \to
e'(k_2)+p_2+\gamma(q_2)$, is shown in Fig. \ref{dvcs}, where
$e(k_1),\ \ e'(k_2)$ denote, respectively, the initial and final
electrons of momenta $k_1,\ k_2$,  and $p_1,\ p_2$ denote the
initial and final momenta of the target correspondingly.

DVCS is the hard electro-production of a real photon, i.e.
$\gamma^*N \to \gamma N'$. Being a process involving a single
hadron, it is one of the cleanest tools to construct generalized
parton distributions (GPD) \cite{GPD,R3,R4,R6,R7}, which reduce to
ordinary parton distributions in the forward direction. The
theoretical efforts and achievements are supported by the
experimental results from HERMES, HERA  and CLAS Collaborations,
and encouraging future plans.

DVCS is characterized by three independent four-momenta:
$p=p_1+p_2,\ \Delta=p_2-p_1,$ and $q=(q_1+q_2)/2,$ where the
vectors $p_1\ (q_1)$ and $p_2\ (q_2)$ refer to the incoming and
outgoing proton (photon) momentum, respectively. Most of the
papers on deep inelastic scattering (DIS) and DVCS are based on
the operator product expansion with extensive use of the
light-front variables. Otherwise, the conventional Bjorken
variable is $x=\frac{Q^2}{2p_1\cdot q_1}, \ Q^2=-q_1^2$, and
$\xi=-\frac{q^2}{q\cdot P}$ is the generalized Bjorken variable.
If both photons were virtual, we would have an extra scaling
variable $\eta=\frac{\Delta\cdot q}{p\cdot q},$ the skewedness (or
skewness) \cite{DVCS,Pire}. The reality of the outgoing photon
implies the presence of only one scaling variable, namely, for
$q_2^2=0$ one has
\begin{equation}
\eta=-\xi\Bigl(1+\frac{\Delta^2}{2Q^2}\Bigr)^{-1}.
\end{equation}
The generalized and ordinary Bjorken variables are related by
\begin{equation}
\xi=x\frac{1+\frac{\Delta^2}{2Q^2}}{2-x+x\frac{\Delta^2}{Q^2}}.
\end{equation}

Our starting point is a complex scattering amplitude depending on
three variables, $\xi,t$ and $Q^2$,  defined by Fig. 1 and the
corresponding legend. Even though our paper is devoted to DIS of
Fig. 2 and relevant CLAS data, we bear in mind the close relation
between DVCS and DIS, the latter being the limiting case of the
former.

Most of the papers on this subject are based on the factorization
properties, separating the perturbative and non-perturbative
dynamics (``handbag'' diagram), according to which, at large
$Q^2$, lowest order perturbation decouples from hadronic dynamics
during the short time of interaction. While factorization in hard
scattering processes is valid to all orders in perturbation
theory, a considerable fraction of the existing data comes from
so-called soft region of small and intermediate values of $Q^2$
($Q^2 \sim 1$ GeV$^2$), where the present non-perturbative
approach can be compared with the relevant successful pQCD
calculations \cite{R1,R2}. Although $t-$ dependence at small $t$
is outside the perturbative QCD (pQCD) domain, nontrivial forms of
the $t-$ dependence at a proper scale suggested recently
\cite{R3,R4,R5} can be confronted to those following from
Regge-dual models.

The phase of the DVCS amplitude experimentally is extracted from
the interference between the DVCS and Bethe-Heitler amplitudes,
like in the case of the Coulomb interference in the forward cone
of elastic hadron scattering. While pQCD factorization details
\cite{R1,R6,R7} how to calculate the real part of the DVCS
amplitude, any Regge-dual model contains the phase explicitly, its
form depending on the available freedom (form of the Regge
singularity, shape of the trajectories etc) inherent in this type
of models. One can hope that the results of the pQCD calculation
will reduce this freedom in the future. Alternatively, this phase
can be approximately reconstructed by means of the dispersion
relations or their simplified version of the derivative dispersion
relations, as it was done in ref. \cite{FFS}.

In a series of papers we initiated the study of DIS and DVCS
within a Regge-dual approach. Its virtue is the presence in the
scattering amplitude of $t-$dependence and of the phase as well as
its explicit energy dependence, compatible with unitarity. At high
energies, the contribution of a dipole pomeron \cite{JMaP}
dominates, while at moderate and low energies subleading
contributions (secondary reggeons) become important. Moreover, by
duality, at low energies, $t-$ channel Regge pole exchanges are
replaced by direct-channel reggeons.

No hard scale factorization is assumed in this approach. External
photons interact with the proton via vector meson (or generalized
vector meson \cite{Schild} ) dominance.

The main idea behind the model is reggeization of the resonances
both in the $s-$ and $t-$ channels. Nonlinear, complex Regge
trajectories replace individual resonance contributions. The
resulting scattering amplitude is a complex function of the
Mandelstam variable $s, t, u$ and of the photon virtuality $Q^2$.
Its imaginary part in the forward direction, $t=0$ corresponds to
ordinary distributions or structure functions (SF), describing
inclusive (e.g. electron-proton) scattering, while the whole
amplitude is directly related to exclusive deeply virtual Compton
scattering and corresponding general parton distributions.

In Refs. \cite{JMP,JMP-1,FFJLM} dual amplitudes with Mandelstam
analyticity (DAMA) were suggested as a model for DVCS or DIS. We
remind that DAMA realizes duality between direct-channel
resonances and high-energy Regge behavior (``Veneziano-duality'').
By introducing $Q^2$-dependence in DAMA, we have extended the
model off mass shell and have shown \cite{JMP,JMP-1} how
parton-hadron (or ``Bloom-Gilman'') duality is realized in this
model. With the above specification, DAMA can serve as explicit
model valid, in principle, at all values of the Mandelstam
variables $s$, $t$ and $u$ as well as for any $Q^2,$ thus
realizing duality "in two dimensions": between hadrons and
partons, on the one hand and between resonances and Regge
behavior, on the other hand. The latter property opens the way of
linking JLab (large x, resonances) and HERA (small x, Regge)
physics.

Recently new data on inclusive  electron-proton cross section in
the resonance region ($W<2.5$ GeV) at momentum transfers $Q^2$
below $4.7$ (GeV/c)$^2,$ measured at the JLab (CEBAF) with the
CLAS detector \cite{Osipenko} were made public. In the present
paper we discuss an analysis of the new CLAS data within this
model.

\begin{figure}[htb]
\begin{center}
\includegraphics[height=6cm,angle=0]{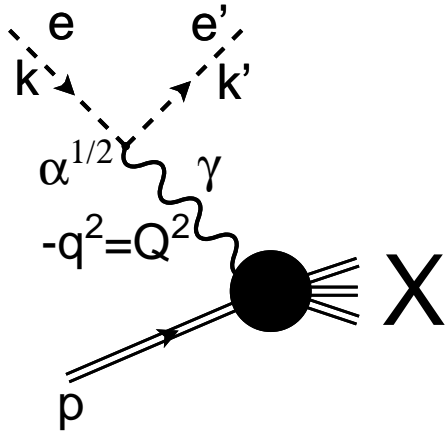}
\end{center}
\caption{Kinematic of deep inelastic scattering.}
\label{rat}
\end{figure}

The kinematics of inclusive electron-nucleon scattering,
applicable to both high energies, typical of HERA, and low
energies as at JLab, is shown in Fig. \ref{rat} (see ref.
\cite{FFJLM} for more details).

Studies of the complex pattern of the nucleon structure function
in the resonance region have a long history  (see, for example
\cite{Stein}). Among dozens of resonances in the $\gamma^* p$
system above the pion-nucleon threshold only a few of them can be
identified more or less unambiguously. Therefore, instead of
identifying each resonance, one considers a few maxima above the
elastic scattering peak, corresponding to some ``effective''
resonance contributions. Recent results  from the JLab
\cite{Nicu,Osipenko} renewed the interest in the subject
 and they call for a more detailed phenomenological analysis of
the data and a better understanding of the underlying dynamics.

The basic idea in our approach is the use the off-mass-shell
continuation of the dual amplitude with nonlinear complex Regge
trajectories. We adopt the two-component picture of strong
interactions, according to which direct-channel resonances are
dual to cross-channel Regge exchanges and the smooth background in
the $s-$channel is dual to the Pomeron exchange in the
$t-$channel. As explained in ref. \cite{JMP}, the background
 in dual model corresponds to pole terms with exotic
trajectories that do not produce any resonance.

\section{Regge-Dual Structure Function}

In the present section we introduce notations, kinematics and the
Regge-dual model. More details on the model can be found in
earlier paper \cite{JMP,JMP-1,FFJLM,dual2}.

So, we study inclusive, inelastic electron-proton scattering,
whose cross section was measured at JLab and used to determine the
unpolarized structure function $F_2(x,Q^2)$ as well as the
Nachtmann  and Cornwall-Norton moments (see e.g. \cite{Roberts}).

The cross section is related to the structure function by \beq
F_2(x,Q^2)={Q^2(1-x)\over{4\pi \alpha (1+4m^2 x^2/{Q^2})}}
\sigma_t^{\gamma^*p}~, \eeq{m23} where the total cross section,
$\sigma_t^{\gamma^*p}$, includes by unitarity all possible
intermediate states allowed by energy and quantum number
conservation, and we follow the norm \beq
\sigma_t^{\gamma^*p}(s)={\cal I}m\  A(s,Q^2)~. \eeq{m22} used in
Refs. \cite{R8,JMP,JMP-1,FFJLM}. The center of mass energy of the
$\gamma^* p$ system, the negative squared photon virtuality $Q^2$
and the Bjorken variable $x$ are related by \beq
s=W^2=Q^2(1-x)/x+m^2~. \eeq{m21}

In the Regge-dual approach with vector meson dominance implied,
Compton scattering can be viewed as an off-mass shell continuation
of a hadronic reaction, dominated in the resonance region by
non-strange ($N$ and $\Delta$) baryonic resonances. The scattering
amplitude can be written as a pole decomposition of the dual
amplitude and factorizes as a product of two vertices (form
factors) times the propagator: \beq \left[A(s,Q^2)\right]_{t=0} =
{\it N} \left\{ \sum_{r,n} {f_r^{2(n-n_r^{min}+1)}(Q^2) \over n -
\alpha_r(s)} + [A(s,Q^2)]_{t=0}^{BG}\right\}~, \eeq{dualampl}
where $\it N$ is an overall normalization coefficient, $r$ runs
over all trajectories allowed by quantum number conservation (in
our case $r=N^*_1,~N^*_2,~\Delta$) while $n$ runs from $n_r^{min}$
(spin of the first resonance) to $n_r^{max}$ (spin of the last
resonance - for more details see next section), and
$[A(s,Q^2)]_{t=0}^{BG}$ is the contribution from the background.
The functions $f_r(Q^2)$ and $\alpha_r(s)$ are respectively form
factors and Regge trajectory corresponding to the $r^{th}-$term.
(For a comparison of the direct-channel, "reggeized" formula
(\ref{dualampl}) with the usual Breit-Wigner expression see
Appendix \ref{ap1}). Note that only for the first resonance at
each trajectory we have squared form factor, while for the
recurrences the powers of form factors are growing, according to
the properties of DAMA \cite{JMP,JMP-1}.

\subsection{Regge Trajectories}

Any systematic account for the large number of direct-channel
resonances (over 20) contributing to the $\gamma^* N$ total cross
section with different weights is not an easy task. However, this
problem can be overcome with the use of (s-channel) Regge
trajectories, including all possible intermediate states in the
resonance region appearing as recurrences on the trajectories. In
this approach, Regge trajectories play the role of dynamical
variables and the parameters of the trajectories can be fitted
either to the masses and widths of the known resonances or to the
data on DIS cross sections (structure functions), reflecting
adequately the position of the peaks in the SF (or cross sections)
formed by the interplay of different resonances.

The form of the Regge trajectories is constrained by analyticity,
requiring the presence of threshold singularities, and by their
asymptotic behaviour imposing an upper bound on their real part.
Explicit models of Regge trajectories realizing these requirement
were studied in a number of papers \cite{ReggeTraj}. For our
present goals (small- and intermediate energies) a particularly
simple model based on a sum of square root thresholds will be
suitable: \beq \alpha(s)=\alpha_0+\alpha_1 s+\alpha_2(\sqrt
{s_0}-\sqrt{s_0-s}), \eeq{m18} where the lightest threshold,
$s_0$, produces the imaginary part and the heaves thresholds
producing the real part can be approximated here by a linear term.
In our case \cite{JMP,JMP-1,FFJLM} $s_0=(m_{\pi}+m_p)^2$.

For asymptotic, large $s-$ or $t-$ the trajectories turn down to a
logarithm, producing wide angle scaling behavior with a link to
the quark model. This interesting regime, discussed e.g. in ref.
\cite{BJC,JMPac}, however is far away from the resonance region
and will not be included in the present analyses.

In $\gamma^* p$ scattering, mainly the two $N^*$s (isospin 1/2)
and one $\Delta$ (isospin 3/2) resonances contribute in the
$s-$channel and thus we will limit ourselves to considering these
three terms, plus an additional terms which describe the
background, to be be discussed later.

\subsection{Form Factors}
\label{ffacts}

In our previous work \cite{FFJLM}, we concentrated our attention
on the analytic structure of the scattering amplitudes using a
simple dipole model for the form factors. However, in order to
properly describe the structure function in the resonance region,
it is essential to account for the helicity structure of the
amplitudes. Below we do so following Davidovsky and Struminsky
\cite{DS}, who provided for relevant amplitudes by using the
Breit-Wigner resonance model. The relation between the
Breit-Wigner and the "reggeized" resonance model, to be used can
be found in Appendix \ref{ap1}.

The form factors can be written as a sum of three terms
\cite{BW,CM86,CM98,DS}, $G_+(Q^2)$, $G_0(Q^2)$ and $G_-(Q^2)$,
corresponding to $\gamma^*N\rightarrow R$ helicity transition
amplitudes in the rest frame of the resonance $R$: \beq
G_{\lambda_\gamma} =
{<R,\lambda_R=\lambda_N-\lambda_{\gamma}|J(0)|N,\lambda_N>\over
m}~, \eeq{m16} where $\lambda_R$, $\lambda_N$ and $\lambda_\gamma$
are the resonance, nucleon and photon helicities, $J(0)$ is the
current operator; $\lambda_\gamma$ takes the values $-1, 0$ and
$+1.$ Correspondingly, the squared form factor is given by a sum
\cite{BW,CM86,CM98,DS} \beq
|G_+(Q^2)|^2+2|G_0(Q^2)|^2+|G_-(Q^2)|^2. \eeq{m15}

The explicit form of these form factors is known only near their
thresholds $|\vec q |\rightarrow 0$, while their large-$Q^2$
behavior is constrained by the quark counting rules.

According to \cite{BW}, one has near the threshold
\beq
G_{\pm}(Q^2) \sim|\vec q|^{J-3/2}, \ \ G_0(Q^2)
\sim{q_0\over{|\vec q|}}|\vec q|^{J-1/2}
\eeq{thr1}
for the so-called normal ($1/2^+\rightarrow 3/2^-, 5/2^+, 7/2^-,...$)
transitions and \beq G_{\pm}(Q^2 \sim|\vec q|^{J-1/2}, \ \
G_0(Q^2) \sim{q_0\over{|\vec q|}}|\vec q|^{J+1/2} \eeq{thr2} for
the anomalous ($1/2^+\rightarrow 1/2^-,$ $3/2^+, 5/2^-,...$)
transitions, where \beq |\vec
q|={\sqrt{(M^2-m^2-Q^2)^2+4M^2Q^2}\over{2M}}\,, \eeq{m12a} \beq
q_0={M^2-m^2-Q^2\over {2M}}, \eeq{m12} $M$ is a resonance mass.

Following the quark counting rules, in refs. \cite{CM98} (for a
recent treatment see \cite {DS}), the large-$Q^2$ behavior of
$G$'s was assumed to be \beq G_+\sim Q^{-3},\ G_0\sim Q^{-4},\
G_-\sim Q^{-5}. \eeq{QCR} Let us note that while this is
reasonable (modulo logarithmic factors) for elastic form factors,
it may not be true any more for inelastic (transition) form
factors. Our  Regge-dual model, eq. (\ref{dualampl}), predicts
that the powers of the form factors increase with increasing
excitation (resonance spin). This discrepancy can be resolved only
experimentally, although a model-independent analysis of the
$Q^2$-dependence for various nuclear excitations is biased by the
(unknown) background.

In ref. \cite{DS} the following expressions for the $G$'s,
combining the above threshold- (\ref{thr1}), (\ref{thr2}) with the
asymptotic behavior (\ref{QCR}), were suggested: \beq |G_{\pm}|^2
= |G_{\pm}(0)|^2 ~{\it q}^{2J-3} ~c^{2J-3}(Q'_0) ~c^{m_{\pm}}(Q_0)\eeq{m10}
\beq |G_0|^2=C^2 {q_0^2 \over{|\vec q|^2}} {\it q}^{2J-1}
c^{2a+m_0}(Q_0) c^{2J-1}(Q'_0) \eeq{m9} for the normal transitions
and \beq |G_{\pm}|^2=|G_{\pm}(0)|^2 {\it q}^{2J-1}
~c^{2J-1}(Q'_0) ~c^{m_{\pm}}(Q_0) \eeq{m8} \beq |G_0|^2 = C^2
\left({q_0^2\over{|\vec q|^2}}\right)^{2J-1} c^{2a+m_0}(Q_0)
c^{2J+1} (Q'_0)~, \eeq{m7} for the anomalous ones, where $m_+=3$,
$m_0=4$, $m_-=5$ count the quarks, $C$ and $a$ are free
parameters. For notational convenience we have introduced the
functions \bea {\it q} &=& {|\vec q | \over |\vec q |_{Q=0}
}~,\nonumber \\ c(z) &=& {z^2\over{Q^2+z^2}}~. \nonumber \eea The
form factors at $Q^2=0$ are related to the helicity
photoproduction amplitudes $A_{1/2}$ and $A_{3/2}$ by \beq
|G_{+,-}(0)|=\frac{1}{\sqrt{4\pi\alpha}}\sqrt{M\over{M-m}}|A_{1/2,3/2}|~.
\eeq{m6}

\subsection{The Background}

Apart from the resonances, lying on the $N^*$'s and $\Delta$
$s-$channel trajectories, dual to an effective bosonic \linebreak
($f$-)trajectory in the $t-$channel, one has to consider the
contribution from a smooth background. Following our previous
arguments, \cite{JMP,JMP-1,FFJLM,dual2}, we model it by
non-resonance pole terms with exotic trajectories, dual to the
Pomeron. \beq \left[ A(s,Q^2) \right]_{BG} = \sum_{b=E,E'} G_b
{c^4(Q_b) \over n_b - \alpha_b(s)}\,. \eeq{background} with dipole
form factors, given by $c^2(Q_b)$. The exotic trajectories are
chosen in the form \beq \alpha_b(s) =
\alpha_b(0)+\alpha_{1b}(\sqrt{s_0}-\sqrt{s_0-s})~, \eeq{a3} where
the coefficients $\alpha_b(0)$, $\alpha_{1b}$ and the $Q^2_{b}$
are the free parameters. To prevent any physical resonance, they
are constrained in such a way that the the real part of the
trajectory terminates before reaching the first resonance on the
physical sheet. An infinite sequence of poles, saturating duality,
appears on the non-physical sheet in the amplitude; they do not
interfere in the smooth behaviour of the background (for more
details see \cite{BJKSh}).

Anticipating the results of Sec. IV, we notice that fits to the
data prefer a negative contribution from the second term in the
background. Formally this is compatible with alternative models
(e.g. \cite{Nicu, Osipenko}), but needs to be understood also in
the framework of the present Regge-dual approach.

\section{Comparison with other models}

In this section we would like to indicate on the two important
properties of our Regge-dual model, that should, in principle,
discriminate it from alternative models of DIS in the resonance
region.

Looking at eq. (\ref{dualampl}) one can see that contrary to the
models accounting for each resonance separately here resonances on
each Regge trajectory enter with progressively increasing powers
of the form factors. This makes the present model quite different
from the existing approaches \cite{Stein,Nicu,Osipenko,DS}. Notice
that increasing powers of the transition form factors result in
the suppression of the relevant contributions  from the
recurrences with growing spin, thus explaining the gradual
disappearance of higher excitations. Further comparison of the
experimentally measured transition form factors may discriminate
between two approaches. Work in this direction is in progress.

The second important difference comes form the parameterization of
the background. We describe the background by non-resonating pole
terms (the poles appear on the non-physical sheet, see
\cite{BJKSh}) with exotic trajectories and standard dipole form
factors. The background contribution strongly decreases with
increasing $Q^2$, whereas in "standard" parameterizations
\cite{Stein, Nicu, Osipenko, DS} the background is an increasing
function of $Q^2$. Since resulting fits by different models are
almost equally good, it is difficult to discriminate between these
two options. Studies of the $Q^2-$dependence of the ratio between
a resonance contributions and the background (at fix energy or x)
may resolve this ambiguity and help to better disentangle
resonances from the background.

\section{Analysis of the CLAS data}

In this section, we present our fits to the CLAS data on the
nucleon structure function, $F_2(Q^2,s)$ \cite{Osipenko}.

A similar analysis using earlier data \cite{Nicu} was carried out
in our previous paper \cite{FFJLM}. The main point of the model
considered in \cite{FFJLM} was the inclusion of three prominent
resonances, $N^*(1520)$, $N^*(1680)$ and $\Delta(1232)$ plus a
background, dual to the Pomeron exchange. In that approach the
large number of resonances contributing to the $F_2$ with
different wights was effectively accounted for by letting the SF
to depend on effective trajectories, whose parameters were fitted
to the data. This approach was, in a sense, justified ``a
posteriori'': the parameters of the effective trajectories were
found to be close to these fitted to the spectrum of baryon
resonance. Although the main features of the SF in \cite{FFJLM}
were reproduced by the dual model, the quality of the fit was far
from perfect. The reason for the poor agreement could be
threefold: first, in \cite{FFJLM} we made an extra simplification
by neglecting the helicity structure of the amplitudes, and the
form factors were chosen in a simple dipole form. Including the
spin changes the form factors in a non-trivial way and complicates
the $Q^2-$dependence of the SF. The second point is related to the
parameterization of the background: in \cite{FFJLM} the background
was modeled by one term only, underestimating the magnitude of the
SF in some regions. The third important reason is the quality of
the data - the set of points available was not homogeneous
resulting in a non-uniform weight of the fit. To cure this
deficiency, we performed a preselection of the initial data set, a
procedure that potentially may results in ambiguities. The fits
were improved, although still are not perfect.

\begin{figure}[htb]
\begin{center}
\includegraphics[height=8.5cm,angle=-90]{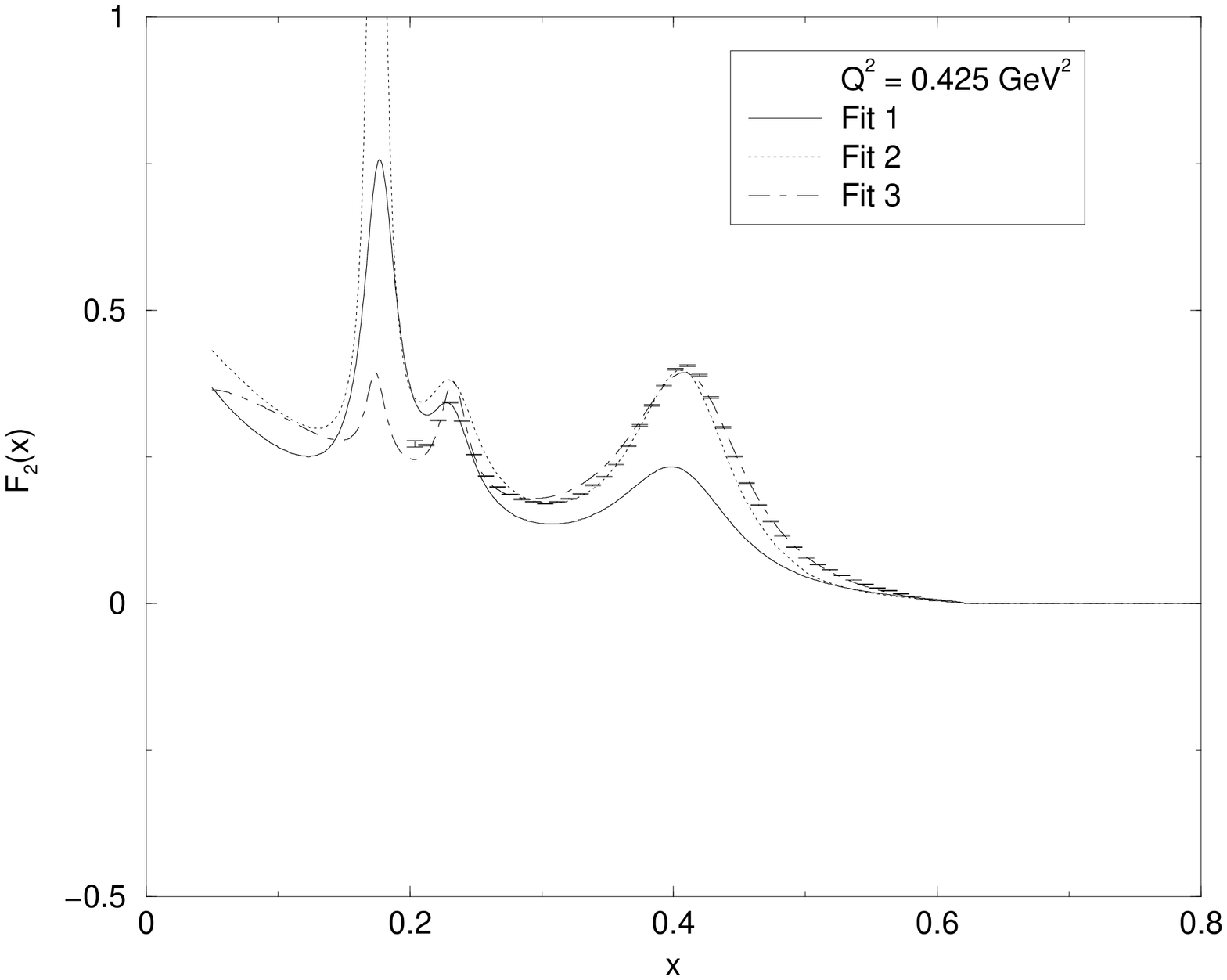}
\includegraphics[height=8.5cm,angle=-90]{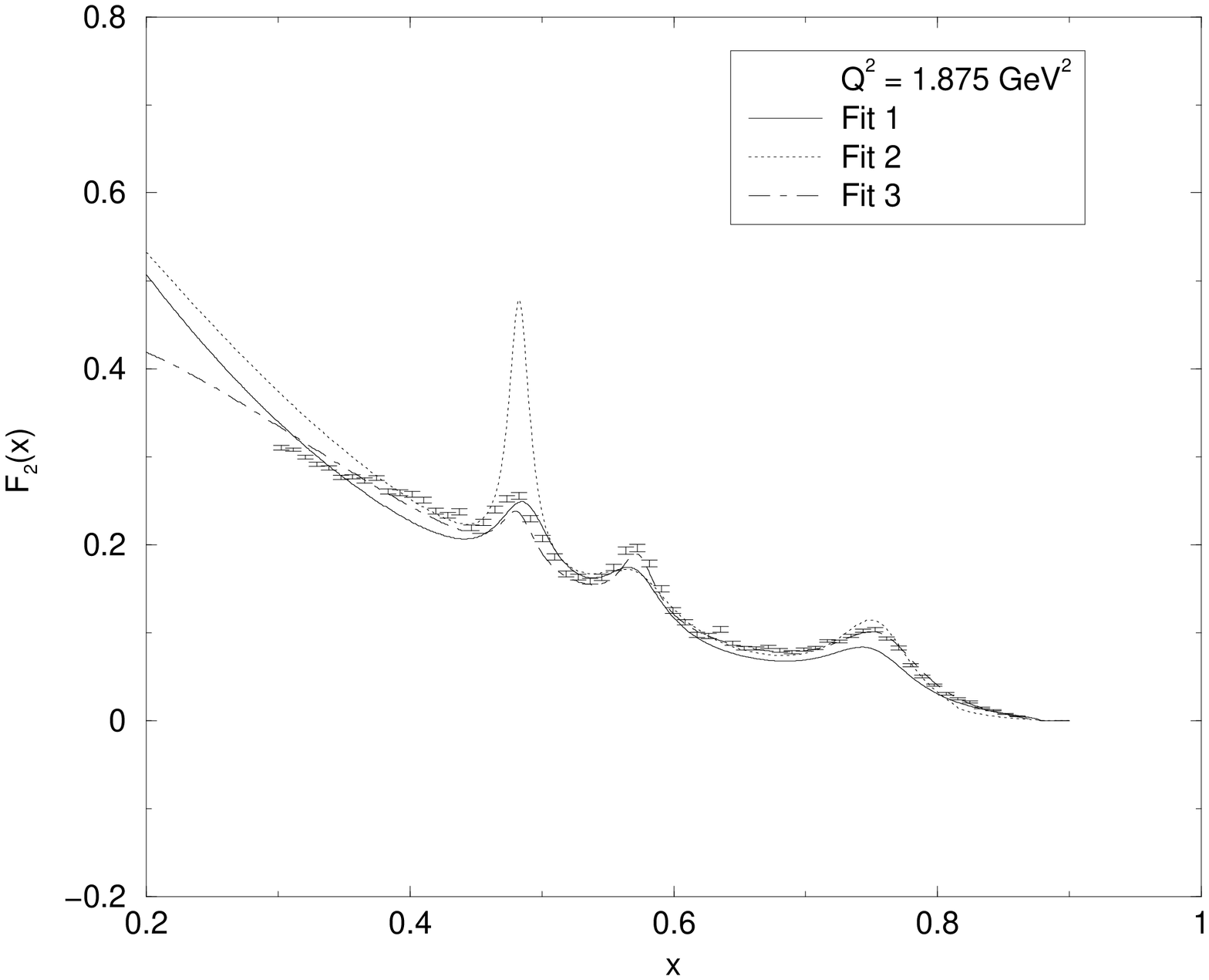}
\includegraphics[height=8.5cm,angle=-90]{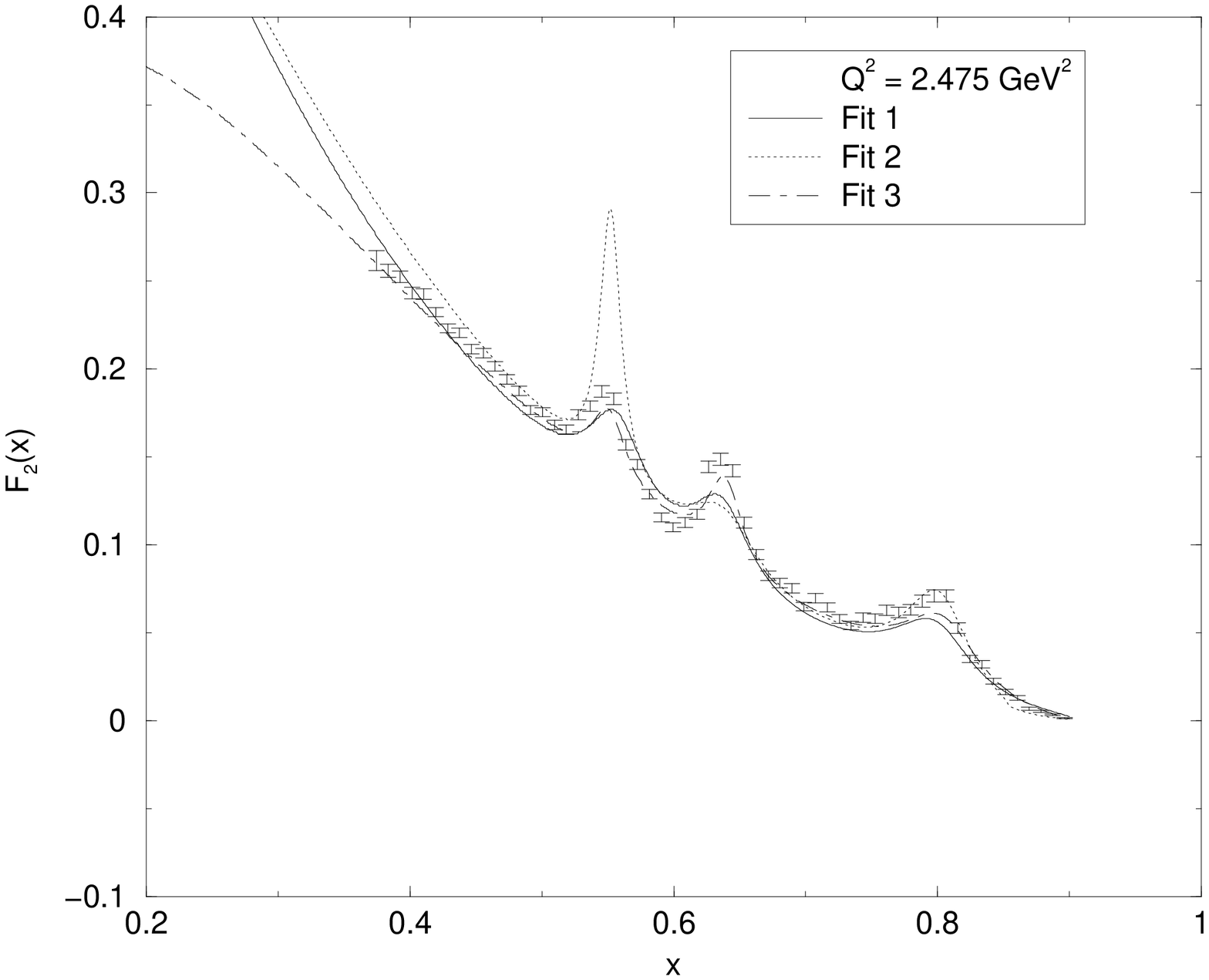}
\includegraphics[height=8.5cm,angle=-90]{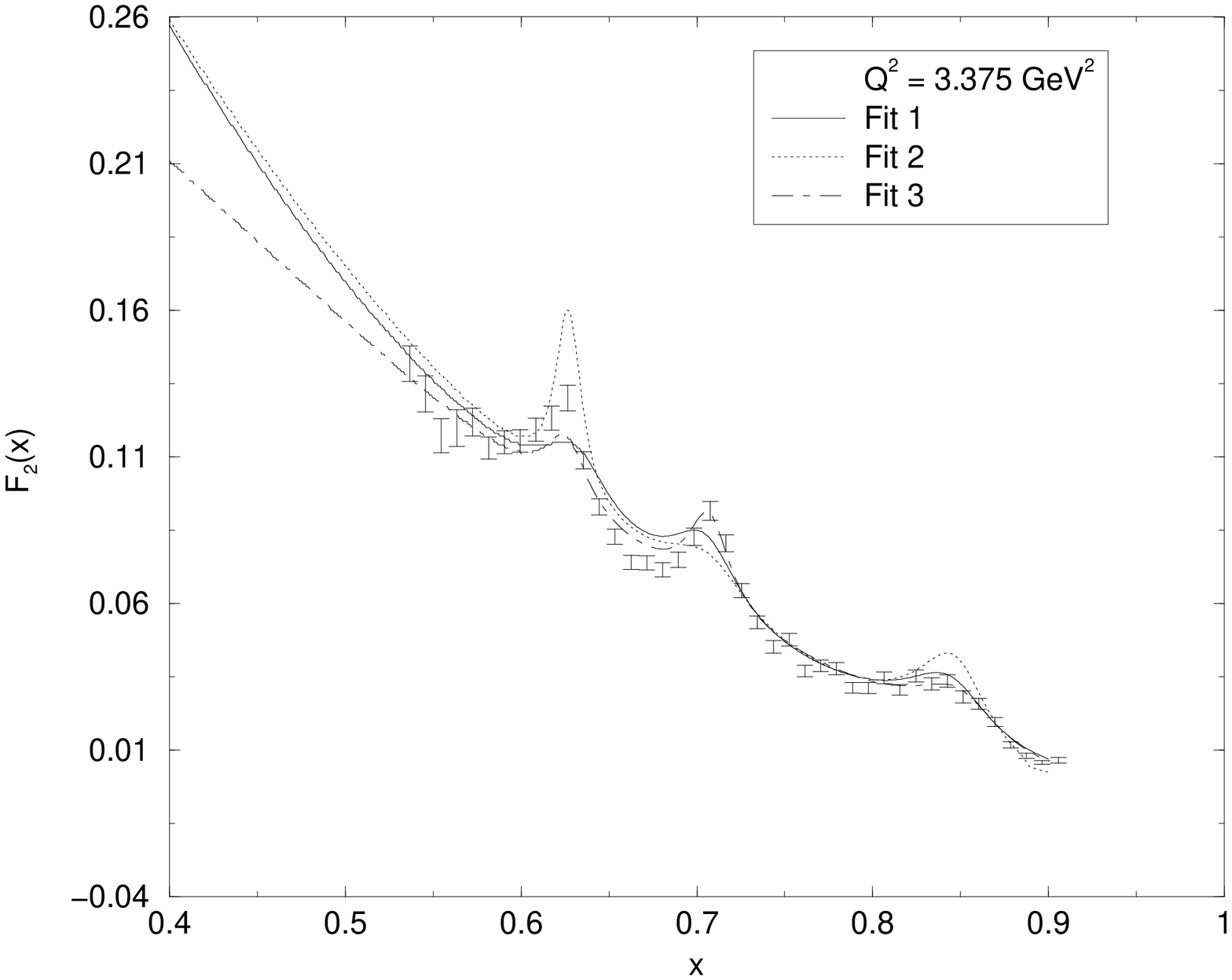}
\end{center}
\caption{Comparison between three different fits performed in the
present model (see text).} \label{diffit}
\end{figure}

Similarly to \cite{FFJLM}, here we also include only the
contribution from three dominant resonances: $N^*(1520)$,
$N^*(1680)$ and $\Delta(1232)$ and we implement this by using
three baryon trajectories with one resonance on each of them. By
considering such resonances as ``effective'' contributions to the
SF, we are able to treat the large number of resonances that
contribute, with different weights, to the SF.

The imaginary part of the scattering amplitude can then be
written, according to (\ref{dualampl}), as a sum of the
contribution from the resonances plus the background, $$ {\cal
I}m\ A(s,Q^2) = {\it N} \left\{[{\cal I}m\  A(s,Q^2)]_{R} + [{\cal
I}m\ A(s,Q^2)]_{BG}\right\}~.$$ Accordingly, the resonance
contribution takes the following form: $$ [{\cal I}m\
A(s,Q^2)]_{R} = \sum_{j=\Delta,~N_1,~N_2} f^2_j(Q^2) {{\cal I}m_j
\over (n_j-{\cal R}e_j)^2+{\cal I}m_j^2}~, $$
 with
${\cal R}e\ $ and ${\cal I}m\ $ denoting the real and imaginary
part of the relevant Regge trajectory, and the form factors are
calculated as described in sec. \ref{ffacts}. For instance, the
form factor for the $\Delta$ resonance can be written as
\begin{equation}
f_{\Delta}^{2}(Q^2) = {\it q}^{2} c^2(Q'_0) \left( c^3(Q_0)
|G_+(0)|^2 + c^5(Q_0) |G_-(0)|^2 \right)\;;
 \label{delta}
\end{equation}
similar expressions can be cast for other contributions.

The imaginary part of the forward scattering amplitude coming from
the background can be easily obtained from (\ref{background}):
\bea [{\cal I}m\  A(s,Q^2)]_{BG}= \sum_{j=E,E'} G_j ~c^4(Q_{j})~
{{\cal I}m_{j} \over (n_{j}^{min}-{\cal R}e_{j})^2+{\cal
I}m_{j}^2}~. \nonumber \eea Here $n_{j}^{min}$ is the lowest
integer, larger than Max $[{\cal R}e_{j}]$, ensuring that no
resonances will appear on the exotic trajectory. The advantage of
such choice is that the two terms of the background depend on two
different scales, $Q^2_{E}$ and $Q^2_{E'}$, so they will dominate
in different regions.

The model constructed in this way, has 23 free parameters: each
resonance is characterized by three (the intercept is kept fixed)
coefficients describing the relevant Regge trajectory plus the two
helicity photoproduction amplitudes (see eq. (\ref{m6})). The form
factors (see sec. \ref{ffacts}) leave only two free parameters,
$Q_0$ and $Q'_0$. Finally, the background, contains 8 free
parameters: 4 for the two exotic trajectories, 2 energy scales
$Q_E$ and $Q_{E'}$ and two amplitudes $G_E$ and $G_{E'}$. With the
overall normalization factor, $\it N$ this gives a total of 23
free parameters.

The resulting fits to the CLAS data, performed by using MINUIT
\cite{minuit}, are presented in Table 1 
and together with the experimental data is shown for various $Q^2$
bins  in Figs. (\ref{labf1}-\ref{labf6}).

\begin{table}[htb]
\begin{tabular}{|c|c|c|c|c|}
\hline
 & parameters & Fit 1 & Fit 2 & Fit 3 \\
\hline
          &  $\alpha_{0}$ & -0.8377$^\diamond$  & -0.8377$^\diamond$   & -0.8377$^\diamond$ \\
          &  $\alpha_{1}$ [GeV$^{-2}$] &  0.9500$^\diamond$    & 0.9402   &  0.9825 \\
$N_1^*$   &  $\alpha_{2}$ [GeV$^{-1}$] &  0.1473$^\diamond$  &
0.1757   & 0.0920
\\
          &  $A^2(1/2)$ [GeV$^{-1}$]  &0.0484E-2$^\diamond$ &0.0484E-2$^\diamond$ &0.8647E-2\\
          &  $A^2(3/2)$ [GeV$^{-1}$]  &0.2789E-1$^\diamond$ &0.2789E-1$^\diamond$&0.9634E-2\\
\hline \hline
          &  $\alpha_{0}$ & -0.3700$^\diamond$   & -0.3700$^\diamond$   & -0.3700$^\diamond$ \\
          &  $\alpha_{1}$ [GeV$^{-2}$]&  0.9500$^\diamond$   & 0.9724  &  0.9551 \\
$N_2^*$   &  $\alpha_{2}$ [GeV$^{-1}$] &  0.1471$^\diamond$ &
0.0575  & 0.0949
\\
          &  $A^2(1/2)$ [GeV$^{-1}$]  &0.0289E-2$^\diamond$ &0.0289E-2$^\diamond$ &0.9724E-2\\
          &  $A^2(3/2)$ [GeV$^{-1}$] &0.1613$^\diamond$ &0.1613$^\diamond$ &  5.1973E-11\\
\hline \hline
          &  $\alpha_{0}$ & 0.0038$^\diamond$  & 0.0038$^\diamond$   & 0.0038$^\diamond$ \\
          &  $\alpha_{1}$ [GeV$^{-2}$] &  0.8500$^\diamond$   & 0.8758   &  0.8605 \\
$\Delta$   &  $\alpha_{2}$ [GeV$^{-1}$] &  0.1969$^\diamond$ &
0.1724  & 0.2005
\\
          &  $A^2(1/2)$ [GeV$^{-1}$] &0.0199$^\diamond$ &0.0199$^\diamond$ & 5.3432E-08\\
          &  $A^2(3/2)$ [GeV$^{-1}$]  &0.0666$^\diamond$ &0.0666$^\diamond$ & 0.0866\\
\hline \hline
          &  G$_{E_1}$    &  6.5488    & 2.8473    & 3.6049 \\
          &  $\alpha_{0}$ &  0.3635     & 0.7014    & 0.3883 \\
          &  $\alpha_{2}$ [GeV$^{-1}$] &  0.1755     & 0.1575    & 0.3246 \\
$E_1$     &  $Q_{E_1}^2$ [GeV$^{2}$] &  5.2645     & 4.5169    &
3.9774   \\
          &  $s_{E_1}$ [GeV$^{2}$]   &  1.14$^\diamond$       & 1.3038    & 1.14$^\diamond$   \\
\hline \hline
          &  G$_{E_2}$    &  ---        & ---    & -0.6520  \\
          &  $\alpha_{0}$ &  ---        & ---    & -0.8929   \\
$E_{2}$   &  $\alpha_{2}$ [GeV$^{-1}$] &  ---        & ---    & 1.7729   \\
          &  $Q_{E_2}^2$ [GeV$^{2}$]   &  ---        & ---    & 2.4634   \\
          &  $s_{E_2}$ [GeV$^{2}$] & ---       & ---    & 1.14$^\diamond$   \\
\hline \hline
         &   $s_0$ [GeV$^{2}$]   & 1.14$^\diamond$        & 1.14$^\diamond$    & 1.14$^\diamond$   \\
\hline
          &  $Q^{'2}_{0}$ [GeV$^{2}$] &  0.4089     & 0.4580    & 0.9998   \\
          &  $Q_{0}^2$ [GeV$^{2}$] &  3.1709     & 2.5180    & 1.8926   \\
\hline
          &  ${\it N}$ [GeV$^{-2}$]  &  0.0408     & 0.0655    & 0.0567 \\
\hline
          &  $\chi^2_{d.o.f.}$    &12.92&4.6886&   1.3005\\
\hline
\end{tabular}
\vskip15pt \noindent{Table 1. Parameters of the fits. The symbol
$^\diamond$ refers to the fixed parameters.} \vskip15pt
\end{table}

To start with, we made a fit by keeping some of the parameters
fixed, close to their physical values, particularly those of the
Regge trajectories and of the photoproduction amplitudes. Also, a
single-term background was used. The resulting fit (fit 1) is
shown in Table 1. Next (fit 2) some of the parameters of the Regge
trajectory were varied. Consequently the $\chi^2$ was improved,
although still remaining unsatisfactory. Finally, we let all the
parameters vary (fit 3) with the result reported in the Table. Fit
3 is good, with $\chi_{d.o.f.} =1.30$. It is worth mentioning that
a comparison with a similar fit performed in \cite{dual2} leading
to $\chi_{d.o.f.} =9.4$ needs care, since in \cite{dual2} only one
term in the background was included, the helicity amplitudes were
kept constant and the dataset used included both data from
\cite{Nicu} and \cite{Osipenko}.

To show the progress in the fits, we plot against the experimental
data the structure functions for four different values of $Q^2$
with the parameters from three different fits - see Fig.
\ref{diffit} .

Having fitted the parameters (from now on we will use parameters
of fit 3), we can now proceed to further calculations (moments)
and analyses (duality relations) of the model.

\section{Moments}

We have calculated the moments of the structure functions using
the explicit expressions and parameters fitted in the previous
section.  These moments can be used, in particular, to estimate
the role of the non-perturbative effects (higher twists).

\begin{figure}[htb]
\begin{center}
\includegraphics[height=16cm,angle=-90]{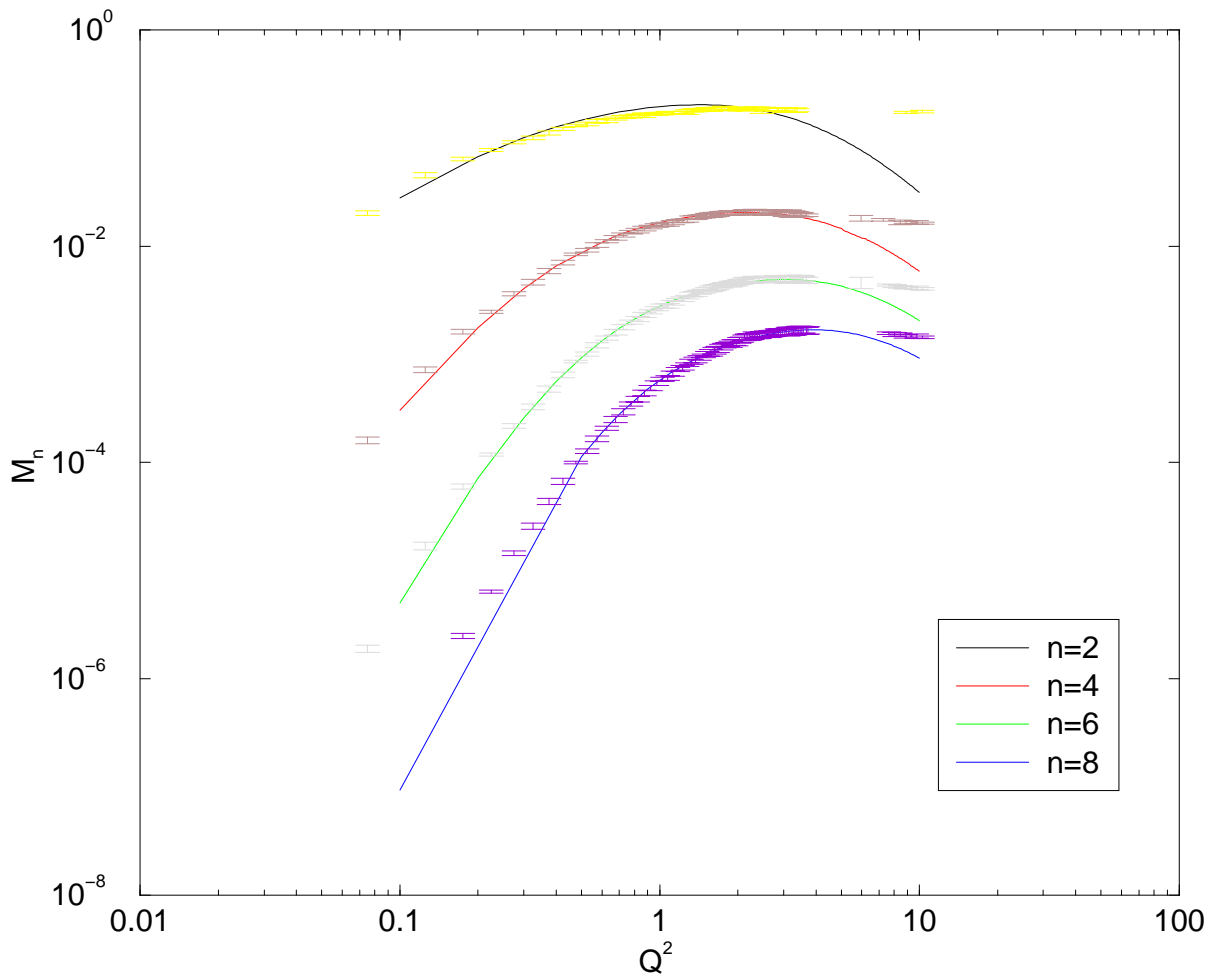}
\end{center}
\caption{Nachtmann moments, $M_n^N$ for $n~=~2,~4,~6,~8$. The plot
compares the moments calculated from the Regge-dual with those
extracted from the data and reported in \cite{Osipenko} (inelastic
part).} \label{nachtosi}
\end{figure}

From the operator product expansion (for a comprehensive review
see e.g. \cite{Roberts}) the moments $M_n(Q^2)$ of $F_2$ are
defined as
 \beq M_n(Q^2) = \sum_{J=2k}^{\infty}
E_{nJ} (\mu, Q^2) O_{nJ}(\mu) \left({\mu^2\over
Q^2}\right)^{(J-2)/2}~, \eeq{a4} where $k=1,2,...$, $\mu$ is a
factorization scale, $O_{nJ}(\mu)$ is the reduced matrix element
of the local operators with definite spin $n$ and twist $J$,
related to the non-perturbative structure of the target, $E_{nJ}
(\mu, Q^2)$ is a dimensionless coefficients related to the small
distance behaviour.

The leading twist term $\tau=2$ is well established in pQCD, while
higher twists are indicators of the non-perturbative and confining
effects. In order to study the higher twists, it is essential to
have a complete knowledge of the $F_2$ covering the entire
$x-$range for each fixed $Q^2$. Higher twists can be well
established only with higher moments ($n~>~2$), meanwhile for
$M_2$ their contribution is small even at $Q^2 \sim 1$ GeV$^2$.
Therefore the most interesting kinematical region lies between $0$
and $5$ GeV$^2$ and large values of $x$, where the higher moments
dominate.
 The JLab data and relevant calculations in \cite{Osipenko} cover most of this region.

\begin{figure}[htb]
\begin{center}
\includegraphics[height=9cm]{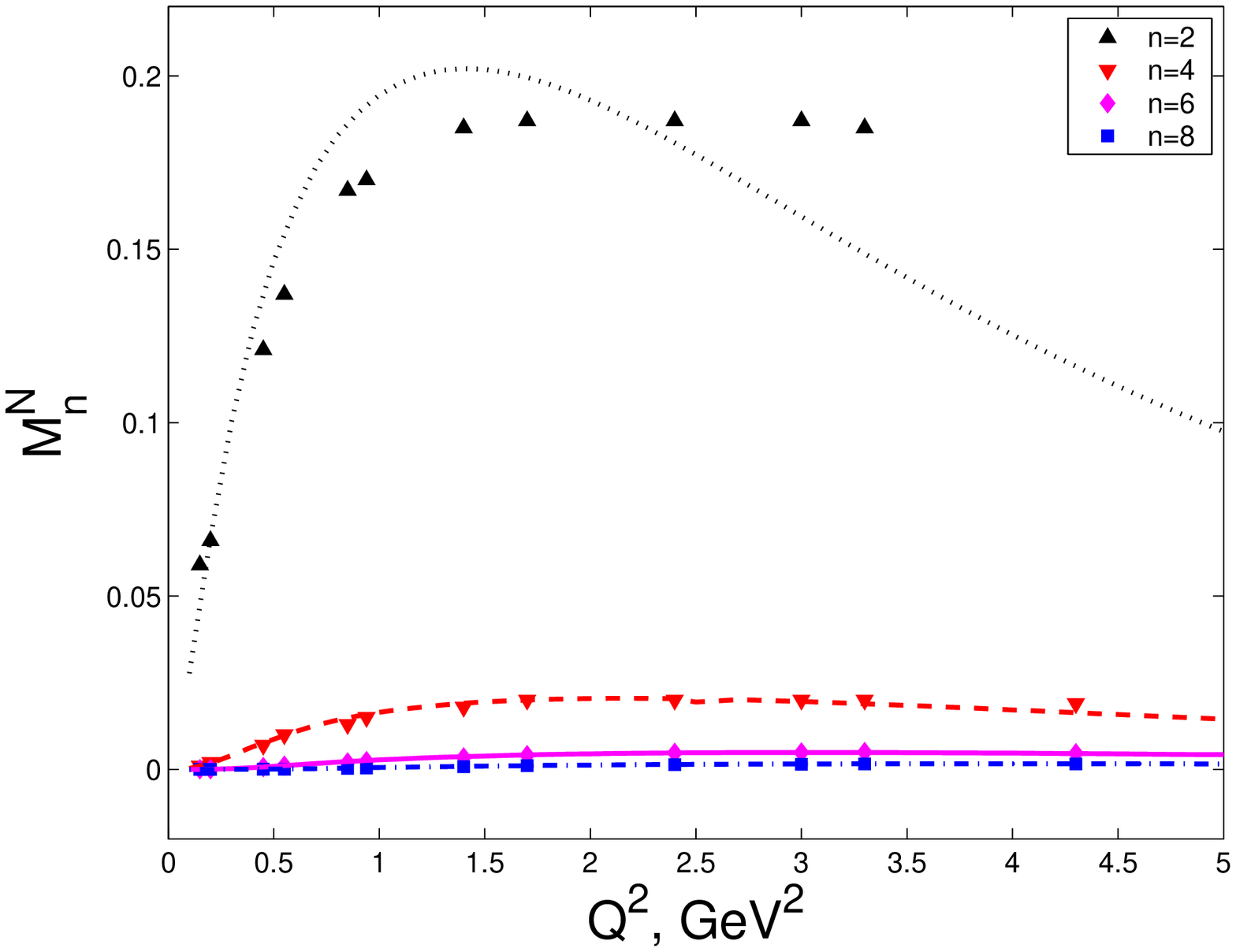}
\includegraphics[height=9cm]{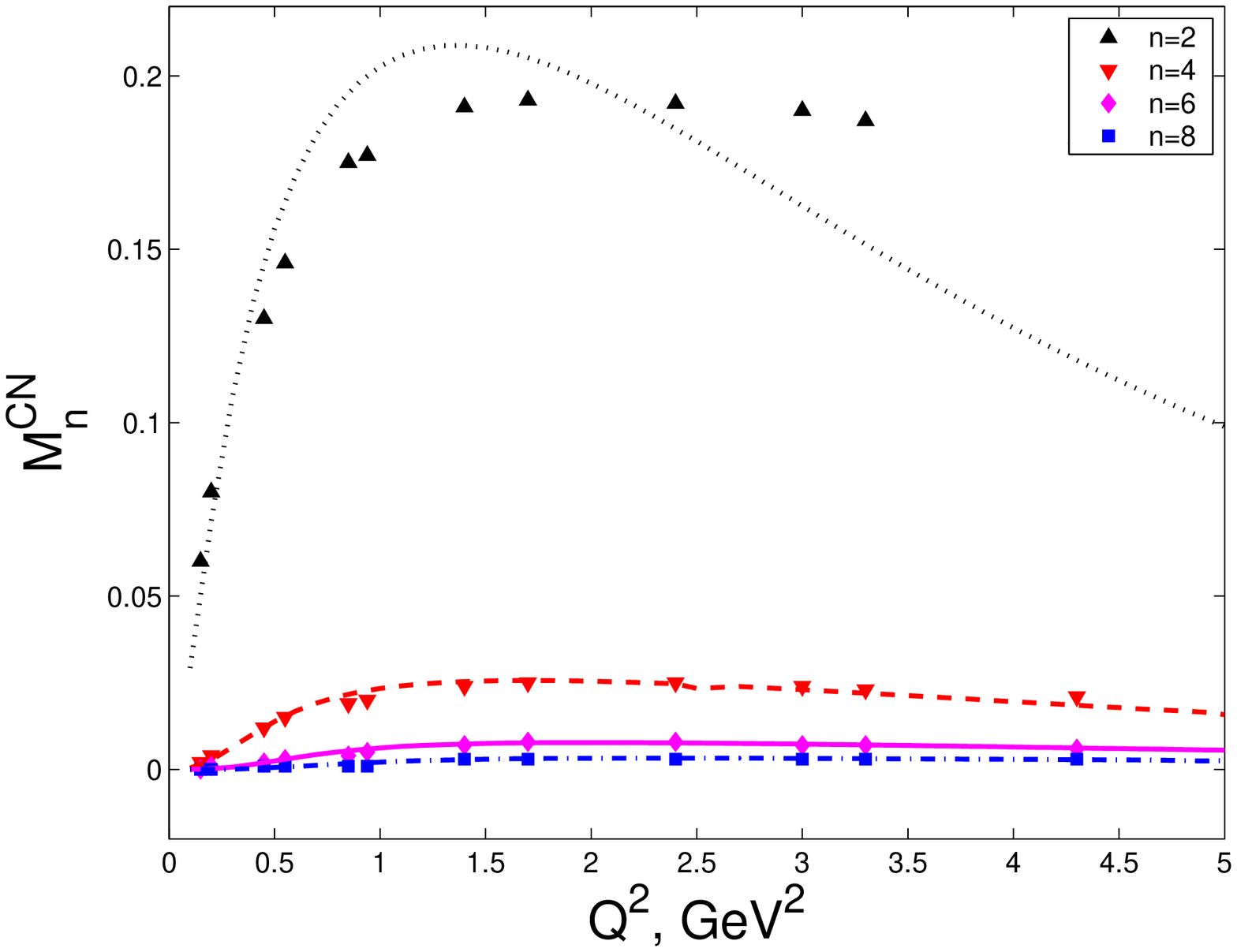}
\end{center}
\caption{Nachtmann moments, $M_n^N$, and Cornwall-Norton moments,
$M_n^{CN}$, for $n~=~2,~4,~6,~8$. These plots show the comparison
between the moments evaluated according to our Regge-dual model
and the values of the moments extracted from the electron-proton
scattering data reported in \cite{armstrong} (inelastic part).}
\label{nachtarm}
\end{figure}

 In the present section we evaluate the Nachtmann (N) and
Cornwall-Norton (CN) moments within our Regge-dual model and
compare them with the data of the CLAS colaboration
\cite{Osipenko} as well as with those from ref. \cite{armstrong}.

The relevant moments are defined as \beq M_n^I(Q^2) = \int_0^1 dx
p_n^I(x) F_2(x,Q^2) \eeq{a5}
 where \bea
\label{mmnts} p^I_n(x) = \left\{
    \begin{minipage}[c]{5cm}
${\xi^{n+1}\over x^3} {\it P}(x,Q^2)$ for I= N \\
$x^{n-2}$,~~~~~~~~~~for I = CN
    \end{minipage}
\right.
\nonumber
\eea
\bea
{\it P}(x,Q^2) &=& \left[{3+3(n+1)r+n(n+2)r^2 \over (n+2)(n+3)}\right]
~,\nonumber\\
r&=&\sqrt{1+4M^2x^2/Q^2}~,\nonumber\\
\xi&=&2x/(1+r)~. \nonumber \eea Please note that in our
calculations the elastic part of the SF (for $x=1$) was not taken
into account (see section III.G in Ref. \cite{Osipenko}).

It is a relatively simple task to obtain the moments by using the
existing numerical integration methods. We have used the
parameters of fit 3 from Table 1. In Fig. \ref{nachtosi} we plot
the Nachtmann moments for $n=2,~4,~6,~8$ together with the results
from \cite{Osipenko}. In Fig. \ref{nachtarm}, the calculated N-
and CN-moments are compared with those from \cite{armstrong}. On
this second set of figures the errors in the momenta are not
displayed; according to \cite{armstrong} they should be less than
5$\%$.

As seen from the figures, the agreement between our model and the
data is quite good in the region $Q^2<5$ GeV$^2$, where the SFs
were fitted to the data. The discrepancies increase with $Q^2$,
away from the measurements.

\section{Duality Ratio}

In this section we check the validity of the parton-hadron duality
for our Regge-dual model by calculating the so-called `duality
ratio' \beq I(Q^2) ={I_{Res} \over I_{Scaling}} \eeq{DR} where
\bea I_{scaling}(Q^2)
&=& \int_{s_{min}}^{s_{max}} ds\ F_2^{scaling}~,\nonumber\\
I_{Res}(Q^2) &=& \int_{s_{min}}^{s_{max}} ds\ F_2^{Res}~,\nonumber
\eea and we have fixed the lower integration limit $s_{min}=s_0,$
varying the upper limit $s_{max}$ equal $5$ GeV $^2$ and $10$
GeV$^2$. These limits imply "global duality", i.e. a relation
averaged over some interval in $s$ (contrary to the so-called
"local duality", assumed to hold at each resonance position). For
fixed $Q^2$ the integration variable can be either $s$ (as in our
case), $x$ or any of its modifications ($x', \  \xi,...$) with
properly scaled integration limits. The difference may be
noticeable at small values of $Q^2$ due to the target mass
corrections (for details see e.g. \cite{Osipenko}).
These effects are typically non-perturbative and, apart from the
choice of the variables, depend on detail of the model.

In choosing the smooth "scaling curve" $F_2^{scaling}$ (actually,
it contains scaling violation, in accord with the DGLAP evolution)
we rely on a model developed in \cite{csernai} and based on a soft
non-perturbative Regge pole input with subsequent evolution in
$Q^2$, calculated \cite{csernai} from the DGLAP equation.

\begin{figure}[htb]
\begin{center}
\includegraphics[height=9cm]{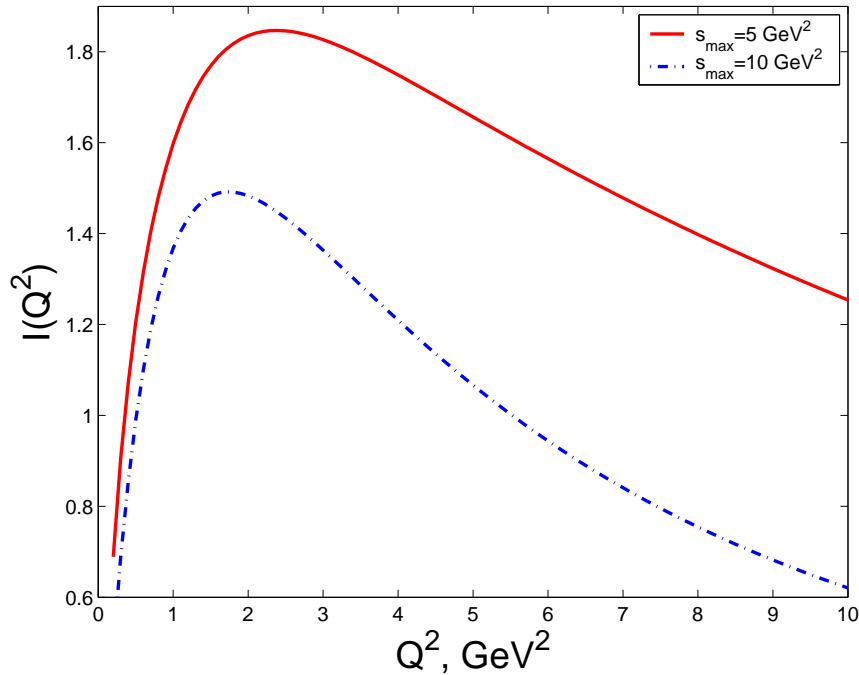}
\end{center}
\caption{Global parton-hadron duality test for different values of
$s_{max}$} \label{phtest}
\end{figure}

\begin{figure}[htb]
\begin{center}
\includegraphics[height=24cm]{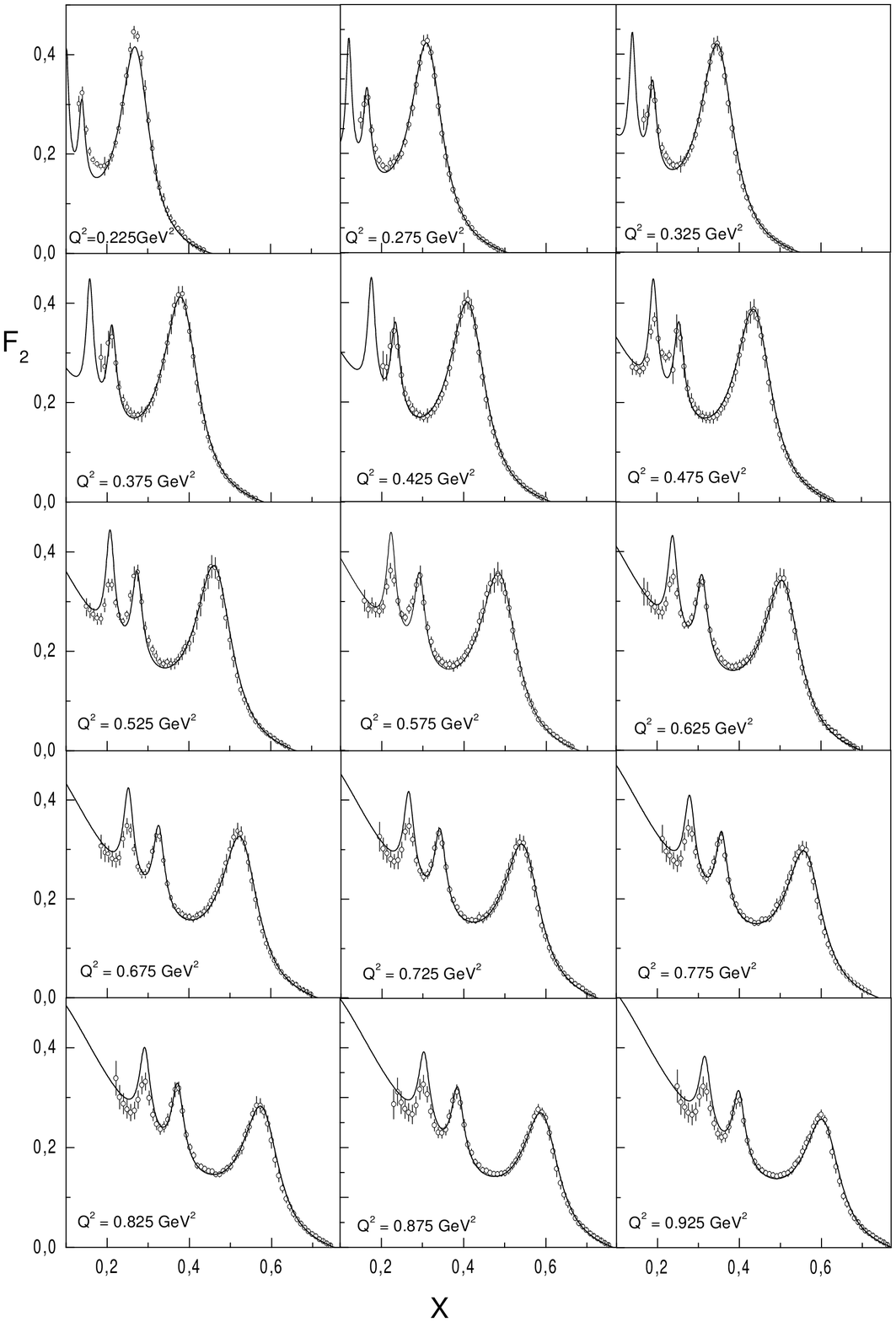}
\end{center}
\caption{Structure function $F_2(x)$ for $Q^2=0.225-0.925$
GeV$^2$. Data are from \cite{Osipenko}, whereas the straight line
is the prediction of our dual model.} \label{labf1}
\end{figure}

\begin{figure}[htb]
\begin{center}
\includegraphics[height=24cm]{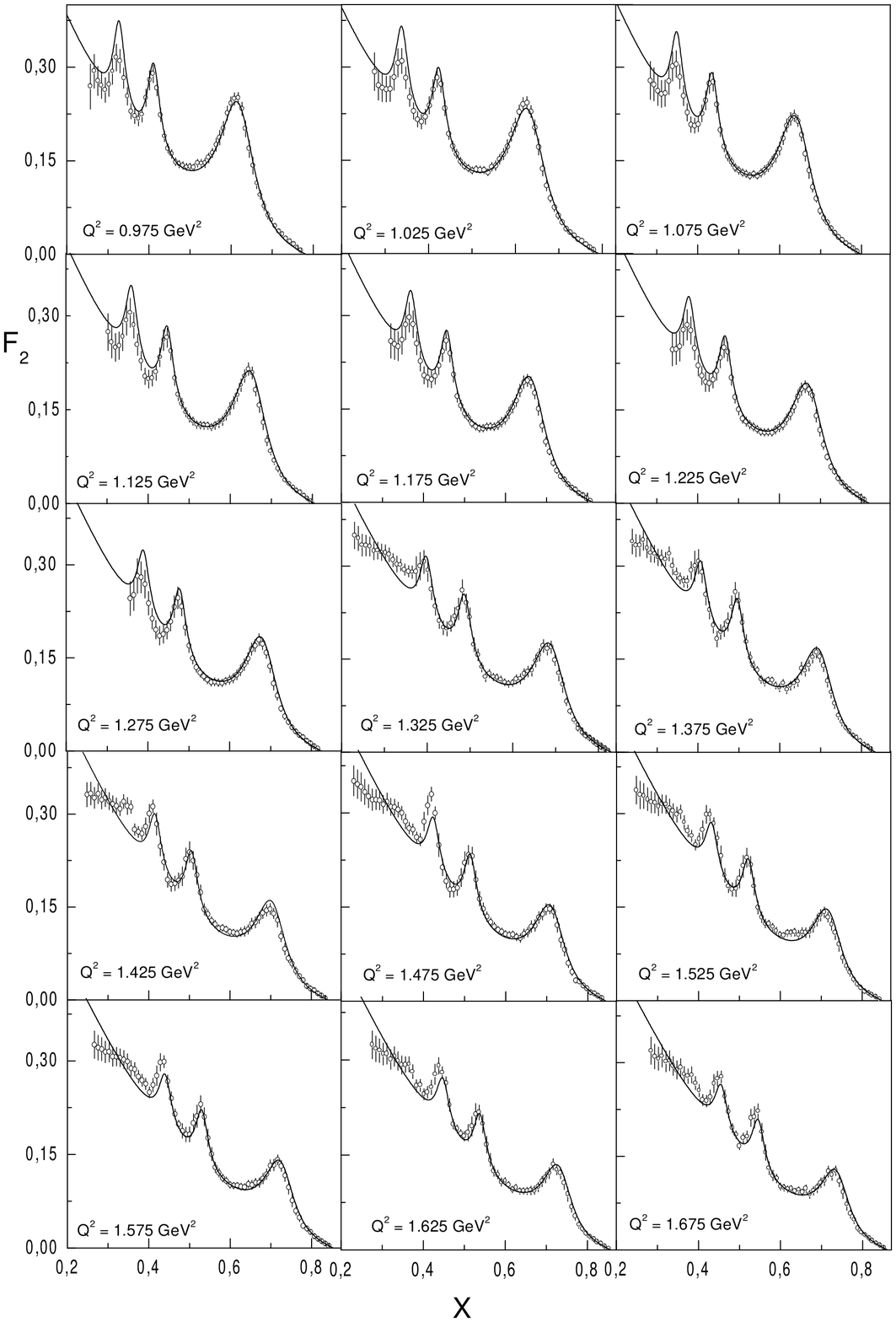}

\end{center}
\caption{Structure function $F_2(x)$ for $Q^2=0.975-1.675$
GeV$^2$.} \label{labf2}
\end{figure}

\begin{figure}[htb]
\begin{center}
\includegraphics[height=24cm]{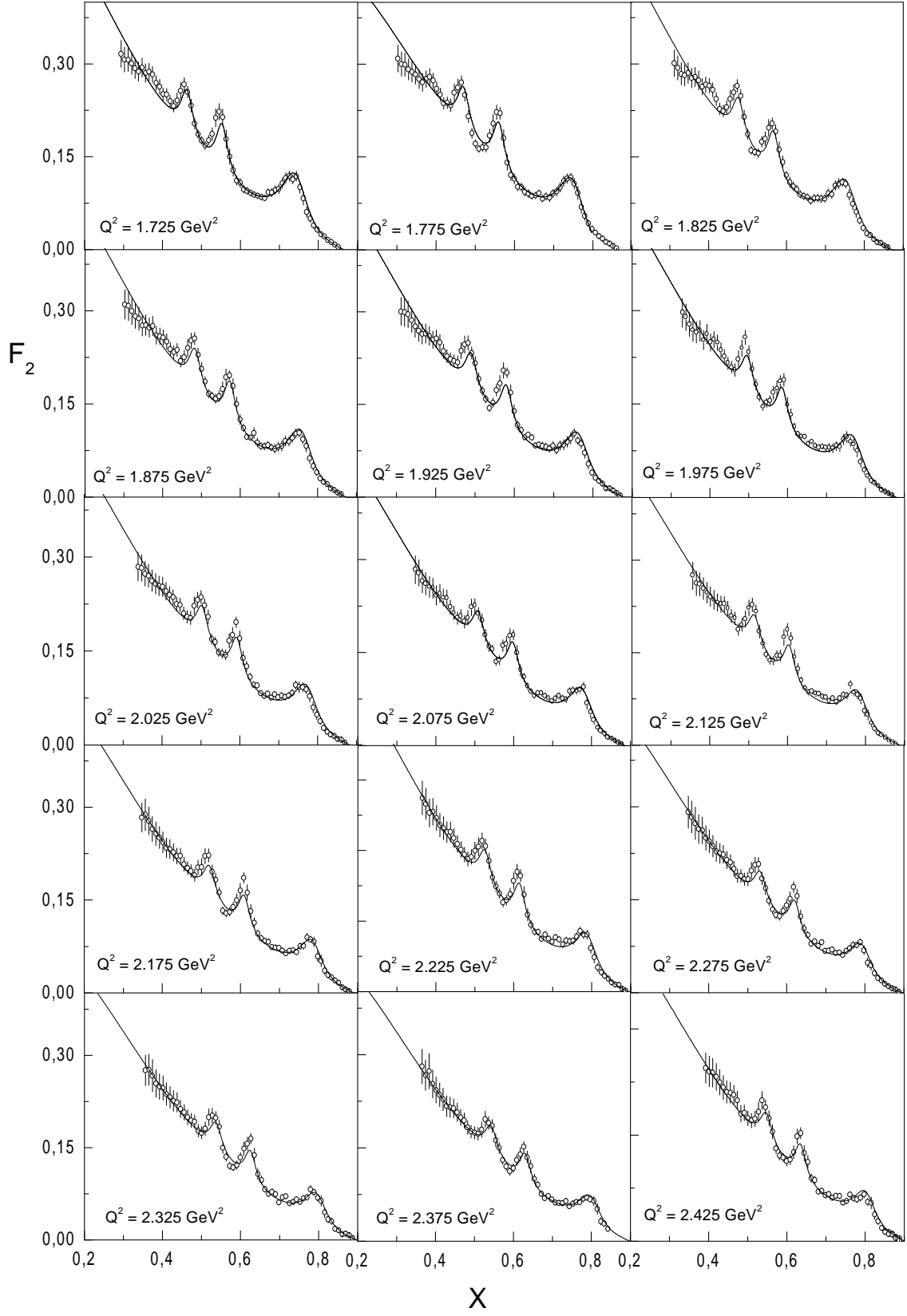}

\end{center}
\caption{Structure function $F_2(x)$ for $Q^2=1.725-2.425$
GeV$^2$.} \label{labf3}
\end{figure}

\begin{figure}[htb]
\begin{center}
\includegraphics[height=24cm]{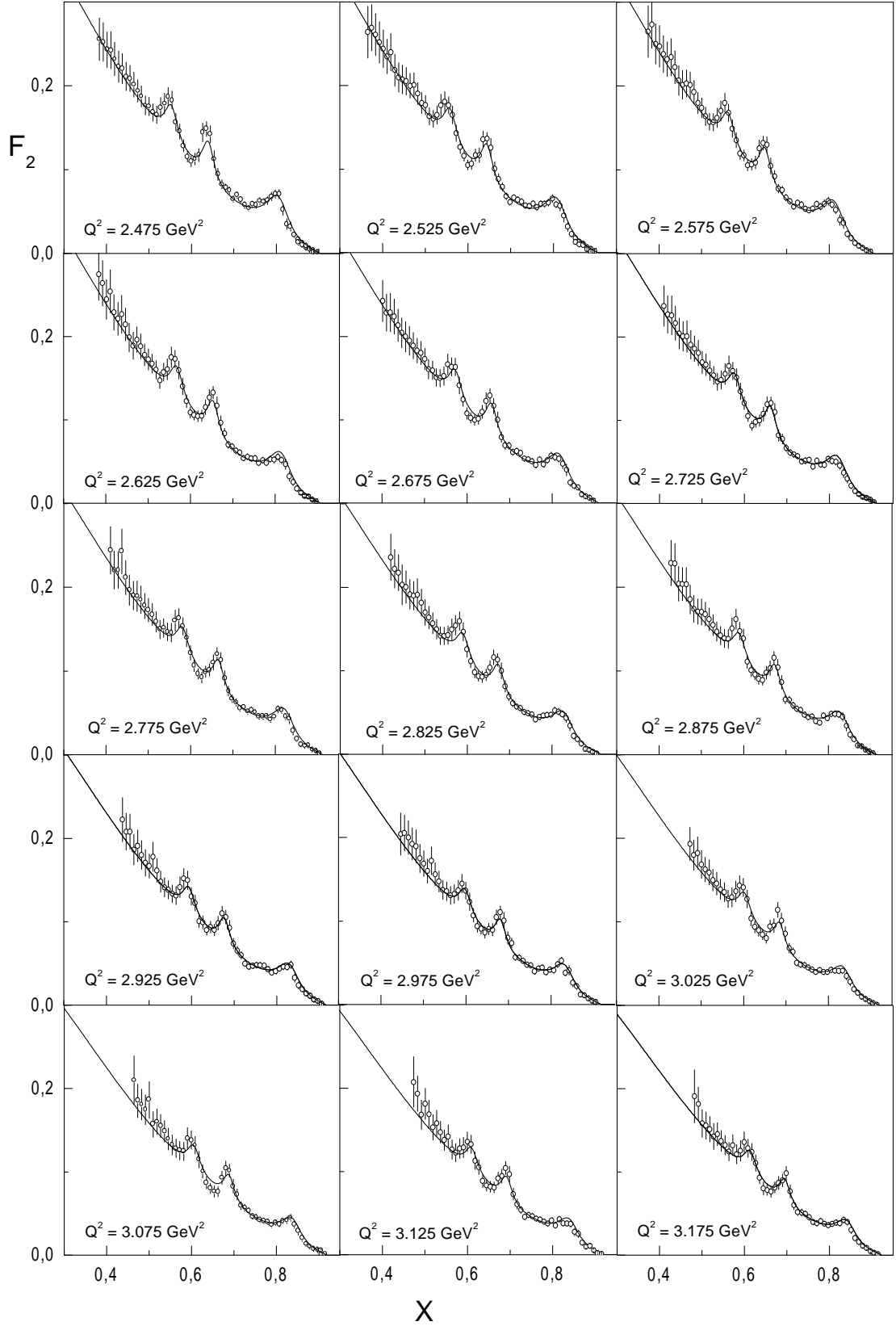}

\end{center}
\caption{Structure function $F_2(x)$ for $Q^2=2.475-3.175$
GeV$^2$.} \label{labf4}
\end{figure}

\begin{figure}[htb]
\begin{center}
\includegraphics[height=24cm]{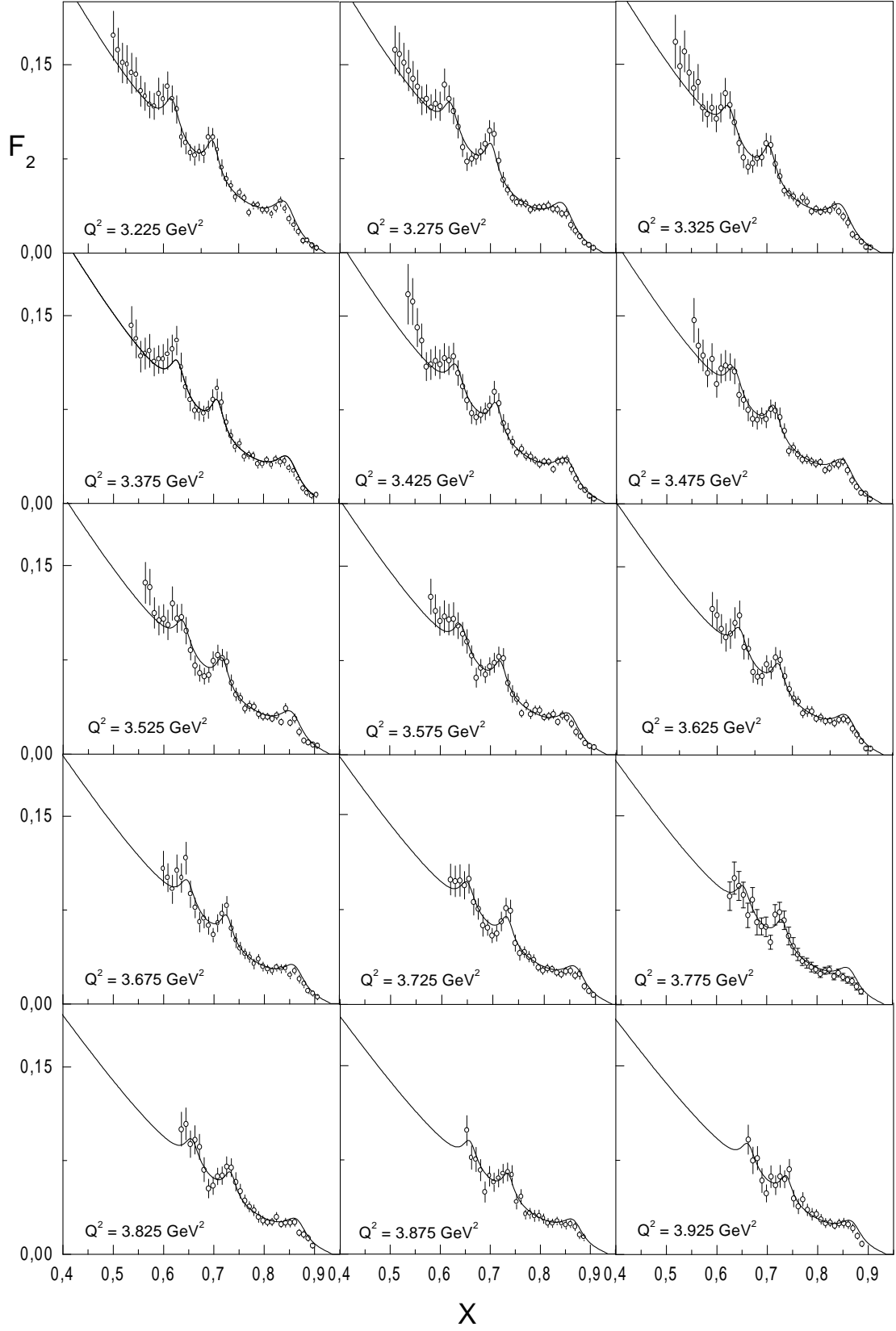}

\end{center}
\caption{Structure function $F_2(x)$ for $Q^2=3.225-3.925$
GeV$^2$.} \label{labf5}
\end{figure}

\begin{figure}[htb]
\begin{center}
\includegraphics[height=24cm]{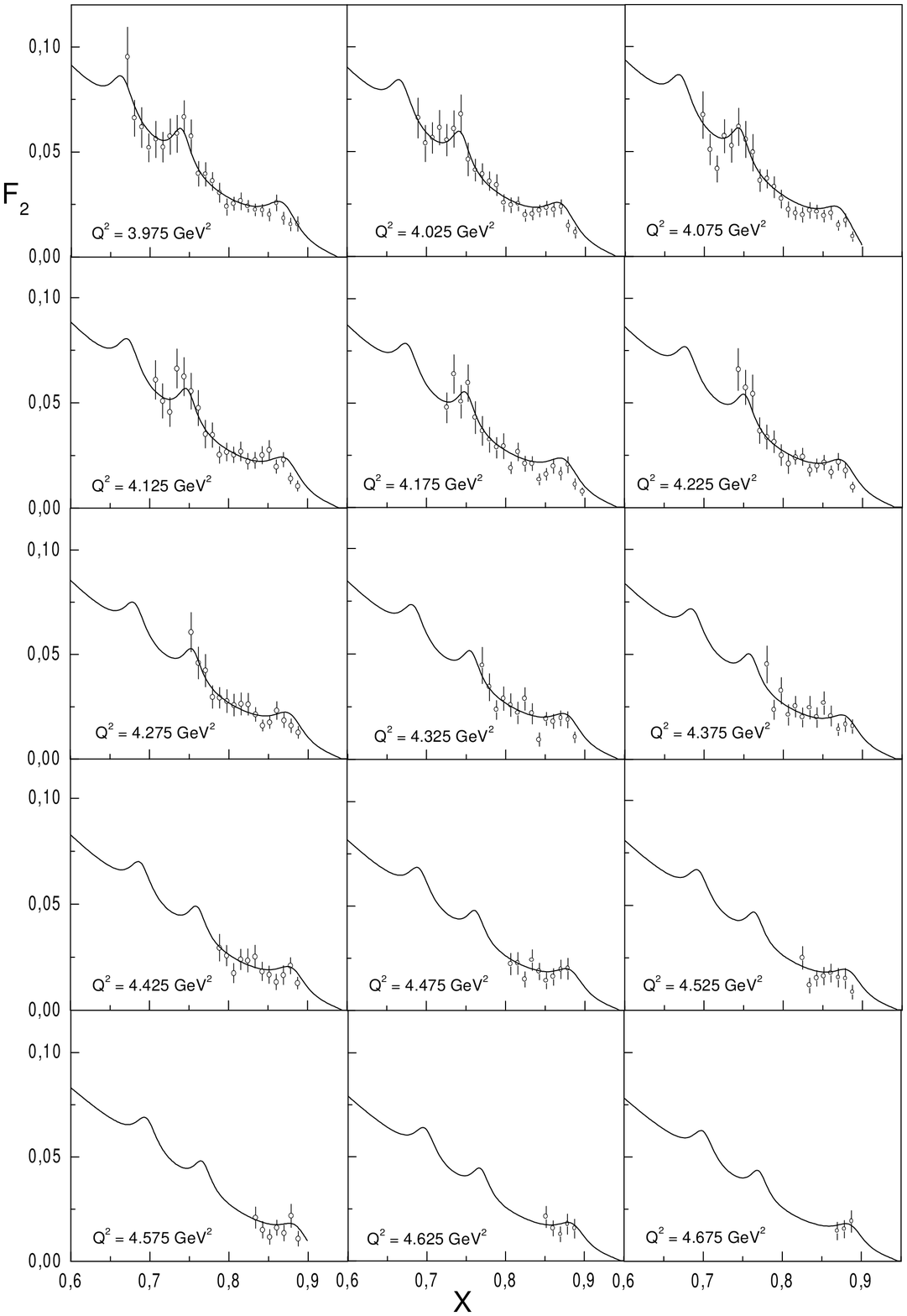}

\end{center}
\caption{Structure function $F_2(x)$ for $Q^2=3.975-4.675$
GeV$^2$.} \label{labf6}
\end{figure}

The function $F_2^{Res}$ is our SF with the parameters of fit 3
(see Table 1). The results of the calculations for different
values of $s_{max}$ are shown in Fig. \ref{phtest}.

Given the available variety and flexibility of the existing
parameterizations for the SFs (see Sec. III) we do no attribute
too much importance to the above duality test. Its validity or
failure to large extent may be caused by accidental interplay of
the details of different parameterizations. By this we do not
intend to raise doubts about the very concept of parton-hadron
duality. Moreover, in our opinion, explicit realization of this
concept, similar to the Veneziano model, should exist and looked
for. Work in this direction is in progress.

\section {Conclusions}\label{s9}

The main objective of the present study is a phenomenological
analyses of the CLAS data in a model within the analytical $S$
matrix approach, complementary to approaches based on pQCD. This
analyses, as well as similar attempts show that achieving good
fits (with low $\chi^2$) to the data is a highly nontrivial task
by itself. The origin of this difficulty is the large number and
high statistics of the data and poor understanding of the
non-perturbative dynamics, typical of the kinematical region where
data are collected.

As repeatedly stressed, our approach does not compete with QCD; it
is aimed to be complementary to QCD in the non-perturbative
domain. The main virtue of our Regge-dual approach is its
generality: potentially, it can be used for any value of its
kinematical variable. From this point of view, of special interest
is the possibility to link low-energy, resonance physics (and the
JLab data) with the high-energy (or low x) physics (from HERA) by
"Veneziano duality" (apart from parton-hadron duality), inherent
in the model.

The price for such generality is the available freedom or
flexibility of the model. It can be, however, further limited by
comparison with other models, pQCD calculations and the data. In
particular,

1. Realistic parameterizations for baryonic trajectories,
satisfying the theoretical constraints yet fitting the data,
should be further elaborated. Work in this direction is in
progress.

2. The separation of resonances from background is
model-dependent. Our parameterization of the background differs
from that introduced long ago (see e.g. \cite{Stein}) and used in
all subsequent papers (e.g. \cite{Nicu,Osipenko}. Its non-orthodox
motivation comes from dual analytical models. At the same time,
fits to the data produce (see Sec. IV) a negative sign in front of
the second term of the background, similar to the "orthodox"
models (e.g. \cite{Nicu,Osipenko}).

3. The present Regge-dual approach generalizes the concept of
transition form factors, continuous in spin. Moreover, higher spin
resonance excitations are accompanied by higher powers of the
relevant transition form factor, and since the Regge trajectories
imply an analytic continuation in spin, the same applies for the
transition form factors.

On the whole, the revival of the analytical methods, namely the
study of various Riemann sheets of the scattering amplitude in the
resonance region (for a recent interesting approach along these
lines see \cite{osip}), and its combinations with the parton model
and QCD is a promising new development in the strong interaction
theory, that may shed new light on the confinement problem.

In estimating the predictive power (or flexibility) of the present
model, we notice that the number of the free parameters here (23)
is comparable or smaller to that in similar fits. For example I.
Niculescu \cite{dissNicu} uses 30 fitting parameters. The virtue
of the present Regge-dual approach is the possibility to extend
the model using the same set of the parameters to the small x
domain, treated in Refs. \cite{FJM,JMP,JMP-1}. Matching the
large-x (Jlab) and small-x (HERA) kinematical regions will remove
or at least reduce substantially the number of the free parameters
and the constrain the flexibility of the model. The realization of
this ambitious goal, already discussed in Refs.
\cite{FJM,JMP,JMP-1}, will depend on the right choice of the $Q^2$
dependence or, alternatively, the correct off mass shell
continuation of the dual amplitude. In the present paper $Q^2-$
dependence was introduced in the resonance region via the
transition form factors.

To conclude, let us once more emphasize that the Regge-dual
approach to DIS and DVCS to large extent is complemental to the
conventional one, based on the presence of a hard scale, when $Q$
(or a mass $M$) is large and the amplitude is calculable up to
corrections of $1/Q$ times logarithms of $Q^2$. In this case
hard-scattering factorization can be applied for any $x$, small or
not.

In the standard approach the generalization of DIS structure
functions to the DVCS amplitude can be illustrated \cite{Diehl} by
the following sequence of transitions \beq F_2\sim {\cal I}m\
A(\gamma^*p\rightarrow \gamma^*p)\rightarrow {\cal I}m\
A(\gamma^*p\rightarrow \gamma p)_{t=0}\rightarrow A
(\gamma^*p\rightarrow \gamma p)_{t=0}\rightarrow A
(\gamma^*p\rightarrow \gamma p)\,.
 \eeq{chain2}

In phenomenological approaches, $t-$ dependence usually is
introduced by simply multiplying the forward scattering amplitude
by arbitrary exponential $e^{Bt}$, incompatible with the shrinkage
of the cone. A consistent, non-factorizable form of the $t-$
dependence was discussed and derived within pQCD in a recent
interesting paper by Freund \cite{R5}.

In the Regge-dual approach, on the other hand, the above sequence
can be inverted: on starts with a complex, $t-$ dependent DVCS
amplitude that can be reduced to the DIS structure function $F_2$
by taking its imaginary part, setting $t=0$ and equating the two
photon momenta. This approach does not require the presence of any
hard scale, such as large photon momenta. The external photons are
assumed to couple to the proton by vector dominance (or
generalized vector dominance \cite{Schild}). In this sense this
approach is typically "non-perturbative". Partons (quarks and
gluons) are not present explicitly but rather implicitly, manifest
in the scaling behavior of the amplitude for large $s,t$ and/or
$Q^2$, as well as in the values of the parameters (e.g. quark
counting). The link between the scaling behavior of the analytic
and quark models is a very interesting but still open problem. It
was approached in a number of papers, e.g. in \cite{JMPac}, where
the large angle scaling behavior in a dual model was achieved by
using Regge trajectories with logarithmic asymptotic behavior.

Although ours is a typically "soft" approach, the quark structure,
small-distance effects, etc are also present their due to the use
of nonlinear Regge trajectories. In particular, the asymptotic
logarithmic behavior of these trajectories could mimic hard
scattering, quark counting etc. \cite{SchSch,JMPac}. These effects
are not factorized, as in the standard approach of \cite{DVCS} and
in most of the related papers, but are continuous, i.e. the
transition from "hard" (perturbative) to "soft" (non-perturbative)
dynamics occurs smoothly, according to the properties of dual
analytical models. The correspondence between the "hard" sector of
this dual model and pQCD (or the quark model) (see
e.g.\cite{SchSch,BJC}) is an interesting problem, meriting further
studies.

\section{Acknowledgments}
We thank M. Osipenko for a useful correspondence. L.J. and V. M.
acknowledge the support by INTAS under Grant 00-00366. The work of
A.F. is supported by the European Community via the award of a
`Marie Curie' postdoctoral Fellowship. A.F. acknowledges the
support of FEDER under project FPA 2002-00748. 
This work has been partially supported by the Ministero
dell'Istruzione dell'Universita' e della Ricerca.

\appendix
\section{Pole decomposition of the dual amplitude and the Breit-Wigner formula}
\label{ap1}

In the vicinity of a resonance, the nucleon structure function can
be can be written in a factorized form \cite{CM98}:
\begin{equation}\label{f2a}
F_2(x,Q^2)=\frac{m\nu Q^2}{\nu^2+Q^2}\delta(W^2-M^2)\,\times
P_F(Q^2)\,
\end{equation}
where  $P_F(Q^2)$ stands for some power of the nucleon
(transition) form factor: this power is two in the standard
approach, as e.g. in ref. \cite{Stein,Nicu,Osipenko,DS}, but
varies (rises) with the resonances spin in the present Regge-dual
approach; $\nu = {p\, \cdot q\over m}=\frac{Q^2}{2mx}$ ($p$ is the
four-dimensional momentum of the nucleon, $q$ is the
four-dimensional momentum of photon, see Fig. \ref{rat}), and $M$
is the mass of the resonance.

This formula determines the contribution of a single, infinitely
narrow resonance to nucleon structure functions. For a wide
resonance, if we replace the delta-function $\delta(W^2-M^2)$ in
the above expression by the familiar Breit-Wigner formula
\begin{equation}\label{resonance}
\frac{1}{\pi}\,\frac{M \Gamma}{(W^2-M^2)^2+M^2 \Gamma^2}\,,
\end{equation}
where $\Gamma$ is the resonance width, what leads to the following
expression \cite{DS}:
\begin{equation}\label{F2_res_fin}
F_2(x,Q^2)=\frac{2m^2 x}{1+4m^2x^2/Q^2}\;\frac{1}{\pi}
\frac{M\Gamma}{(m^2+Q^2(1/x-1)-M^2)^2+M^2\Gamma^2}\times
P_F(Q^2)\,.
\end{equation}

Now let us compare this expression with our eq. (\ref{delta}):
\begin{equation}
F_2(x,Q^2)={Q^2(1-x)\over{4\pi \alpha (1+4m^2 x^2/{Q^2})}}  {\it
N} {{\cal I}m_j \over{ (n_j-{\cal R}e_j)^2+{\cal I}m_j^2}}\times
P_F(Q^2)~, \label{F2_we}
\end{equation}

Expanding the Regge trajectory near a resonance: ${\cal
R}e_j\approx n_j+\{{\cal R}e\ \alpha_j\}^{'}(s-M^2)=n_j+ \{ {\cal
R}e\ \alpha_j\}^{'}(m^2+Q^2(1/x-1)-M^2)$ and introducing the
notation: $\Gamma=\frac{{\cal I}m_j}{\{{\cal R}e\ \alpha_j\}^{'}
M}$, we get the expression:
\begin{equation}
F_2(x,Q^2)={Q^2(1-x)\over{4\pi \alpha (1+4m^2 x^2/{Q^2})}}
\frac{\it N}{ \{{\cal R}e\ \alpha_j\}^{'}}
\frac{M\Gamma}{(m^2+Q^2(1/x-1)-M^2)^2+M^2\Gamma^2}\times
P_F(Q^2)~. \label{F2_we2}
\end{equation}
Notice that $Q^2(1-x)=(s-m^2)x\approx (M^2-m^2)x$ in the vicinity
of the resonance and therefore eqs. (\ref{F2_res_fin}) and
(\ref{F2_we2}) are approximately the same for
\begin{equation}
 {\it N}=\frac{8m^2\alpha \{{\cal R}e\ \alpha_j\}^{'} }{(M^2-m^2)}\,.
\label{norm}
\end{equation}
The obtained value for the normalization coefficient is
approximately (for $M=\sqrt{2}\ m$ and $\{{\cal R}e\
\alpha_j\}^{'}=1$ GeV$^{-2}$) ${\it N} \approx 8 \alpha = 0.058$
GeV$^{-2}$, in agreement with the results of the fit (see Table
1).

\end{document}